\def\call{{\cal L}}

\def\tev{~{\rm TeV}\xspace}
\def\gev{~{\rm GeV}\xspace}
\def\mev{~{\rm MeV}\xspace}

\def\pt{\ensuremath{p_{T}}\xspace}
\def\aeta{\ensuremath{|\eta|}\xspace}

\def\met{\mbox{\ensuremath{\, \slash\kern-.6emE_{T}}}}

\def\nch{\ensuremath{n_{\mathrm{ch}}}\xspace}
\def\sumpt{\ensuremath{\sum{p_{T}}}\xspace}
\def\avgpt{\ensuremath{\langle p_{T} \rangle}\xspace}

\def\Journal#1#2#3#4{{#1} {\bf #2}, #3 (#4)}

\def\NIM{\em Nucl. Instrum. Methods}

\def\PRD{{\em Phys. Rev.} D}
\def\ZPC{{\em Z. Phys.} C}

\documentclass{desyproc}

\usepackage{amsmath,amssymb}
\usepackage{xspace}
\usepackage{hyperref}
\usepackage{multirow}

\begin{document}

% %%%%%%%%%%%%%%%%%%%%%%%%%%%%%%%%%%%%%%%%%%%%%%%%%%%%%%%%%%%%%%%%%%%%%%%%%%%%%%%%
%   Title of your contribution
% %%%%%%%%%%%%%%%%%%%%%%%%%%%%%%%%%%%%%%%%%%%%%%%%%%%%%%%%%%%%%%%%%%%%%%%%%%%%%%%%

\title{Minimum Bias and Underlying Event \\Measurements with ATLAS}

% %%%%%%%%%%%%%%%%%%%%%%%%%%%%%%%%%%%%%%%%%%%%%%%%%%%%%%%%%%%%%%%%%%%%%%%%%%%%%%%%

% %%%%%%%%%%%%%%%%%%%%%%%%%%%%%%%%%%%%%%%%%%%%%%%%%%%%%%%%%%%%%%%%%%%%%%%%%%%%%%%%
%   Authors
% %%%%%%%%%%%%%%%%%%%%%%%%%%%%%%%%%%%%%%%%%%%%%%%%%%%%%%%%%%%%%%%%%%%%%%%%%%%%%%%%

%
%   For the author list please adhere to the format of one of the following
%   three examples:
%
%
%   * use the following for a single author
%
%     \author{{\slshape Joe Smith}\\[1ex]
%     DESY, Notketra{\ss}e 85, 22607 Hamburg, Germany }
%
%   * use the following for several authors
%
%     \author{{\slshape Jean Meunier$^1$, Ruth Miller$^2$,
%     Gerd M\"uller$^3$\footnote{Speaker}, Joe Smith$^3$}\\[1ex]
%     $^1$CERN, 1211 Gen\`eve 23, Switzerland\\
%     $^2$Fermilab, P.O. Box 500, Batavia, IL 60510-0500, USA\\
%     $^3$DESY, Notketra{\ss}e 85, 22607 Hamburg, Germany}
%
%   * use the following for an author speaking on behalf of a collaboration
%
%     \author{{\slshape Joe Smith}  for the FOO Collaboration\\[1ex]
%     CERN, 1211 Gen\`eve 23, Switzerland}
%

\author{{\slshape Michael Leyton}  on behalf of the ATLAS Collaboration\\[1ex]
CERN, 1211 Gen\`eve 23, Switzerland}

% %%%%%%%%%%%%%%%%%%%%%%%%%%%%%%%%%%%%%%%%%%%%%%%%%%%%%%%%%%%%%%%%%%%%%%%%%%%%%%%%

\contribID{ZZ}
\confID{UU}
\desyproc{DESY-PROC-2012-YY}
\acronym{MPI@LHC 2011}
\doi
\maketitle

% %%%%%%%%%%%%%%%%%%%%%%%%%%%%%%%%%%%%%%%%%%%%%%%%%%%%%%%%%%%%%%%%%%%%%%%%%%%%%%%%
%   Abstract
% %%%%%%%%%%%%%%%%%%%%%%%%%%%%%%%%%%%%%%%%%%%%%%%%%%%%%%%%%%%%%%%%%%%%%%%%%%%%%%%%

\begin{abstract}

A summary of some of the recent minimum bias and underlying event measurements by the ATLAS collaboration is given. The results of several analyses using low-luminosity proton-proton collision data from the LHC taken at center-of-mass energies of $\sqrt{s} = 0.9$, 2.36 and 7\tev are presented. Data are compared to predictions by several different Monte Carlo event generators. The measurements expose limitations of the phenomenological models in properly describing the measured observables in all regions of phase space.

\end{abstract}

% %%%%%%%%%%%%%%%%%%%%%%%%%%%%%%%%%%%%%%%%%%%%%%%%%%%%%%%%%%%%%%%%%%%%%%%%%%%%%%%%

% %%%%%%%%%%%%%%%%%%%%%%%%%%%%%%%%%%%%%%%%%%%%%%%%%%%%%%%%%%%%%%%%%%%%%%%%%%%%%%%%
%   Contents
% %%%%%%%%%%%%%%%%%%%%%%%%%%%%%%%%%%%%%%%%%%%%%%%%%%%%%%%%%%%%%%%%%%%%%%%%%%%%%%%%

\section{Introduction}

Properties of Minimum Bias and the Underlying Event have previously been studied over a wide range of energies. In particular, results from experiments at CERN and Fermilab have been used to tune the \textsc{Pythia}~\cite{pythia} and \textsc{Phojet}~\cite{phojet} Monte Carlo (MC) event generators. Due to uncertainties in the modeling of the energy dependence of soft inelastic interactions, these generators give widely varying predictions at LHC energies. Recent results from the ATLAS collaboration are presented here and compared to these predictions. 

\section{LHC and ATLAS}

The Large Hadron Collider (LHC)~\cite{lhc} is a proton-proton ($pp$) collider located at CERN and currently operating at a center-of-mass (CM) energy of $\sqrt{s} = 7\tev$. It is designed to go up to twice that energy, with an instantaneous luminosity of $\call = 10^{34}$cm$^{-2}$s$^{-1}$.

ATLAS~\cite{atlas} is a $4\pi$ general-purpose detector designed for high-luminosity studies at the LHC. Layers of tracking detectors, calorimeters and muon chambers cover almost the entire solid angle around ATLAS. The Inner Detector (ID) is responsible for tracking charged particles within a pseudorapidity range of $\aeta < 2.5$.\footnote{The ATLAS reference system is a right-handed coordinate system with its origin at the nominal interaction point at the center of the detector. Cylindrical coordinates ($r$,$\phi$) are used in the plane transverse to the beam axis, $\phi$ being the azimuthal angle around the beam axis. The pseudorapidity is defined in terms of the polar angle $\theta$ with respect to the beam axis as $\eta = - \mbox{ln tan }(\theta/2)$. The transverse momentum $\pt$ is defined relative to the beam axis.} It consists of a multi-layer silicon tracker with both pixels and strips, in addition to a transition radiation straw tracker, all of which are immersed in a solenoidal magnetic field of 2 T. 

The calorimeters surround the Inner Detector and are responsible for measuring the energies of charged and neutral particles within a pseudorapidity range of $\aeta < 4.9$. The calorimeters are specialized for measuring electromagnetic or hadronic particles, the latter of which include jets of particles formed by hadronization of quarks and gluons. They also detect missing transverse energy (\met) by summing all of the measured energy deposits.

The ATLAS detector has a three-level trigger system. For the measurements presented herein, the first level trigger (L1) relies on the beam pickup timing devices (BPTX) and the minimum bias trigger scintillators (MBTS). The BPTX are composed of electrostatic beam pickups attached to the beam pipe at a distance $z = \pm 175$ m from the center of the ATLAS detector. The MBTS are mounted at each end of the detector in front of the endcap-calorimeter cryostats at $z = \pm 3.56$ m and are segmented into eight sectors in azimuth and two rings in pseudorapidity (2.09 $< \aeta <$ 2.82 and 2.82 $< \aeta <$ 3.84). 

\section{Soft QCD at the LHC}

All physics at the LHC essentially comes from the interactions of quarks and gluons. Hard processes are characterized by high transverse momentum (\pt) and are well described by perturbative QCD. Soft interactions, on the other hand, are characterized by low transverse momentum and require non-perturbative phenomenological models. These soft interactions are actually the dominant processes at the LHC. They can include diffraction, Multiple-Partonic Interactions (MPI), soft initial- and final-state radiation (ISR/FSR), as well as beam-beam remnants. 

While these are all separate phenomena, the different components are often grouped according to experimental trigger. Minimum bias (MB) interactions, for example, are the processes that are selected with a loose trigger intended to select inelastic collisions with as little bias as possible. The Underlying Event (UE) is the collection of all the soft processes that accompany a high-\pt interaction of interest. It is typically studied as a function of the highest-\pt particle in the event. 

Modeling of these soft interactions is important because they impact all other high-\pt measurements. At higher luminosities, for example, minimum bias interactions are a major background, numbering up to 25 interactions on average per bunch crossing at LHC design luminosity. %It is therefore important to have an accurate model of MB for all other high-\pt physics measurements. 
A proper model of the UE is also important for precise high-\pt measurements since it can affect the \met\ resolution, lepton identification and jet resolution. Studying the UE is critical for understanding the evolution of QCD with collision energy, as well as understanding the systematic corrections on many studies such as mass measurements. 

While soft QCD is modeled by some MC generators, including \textsc{Pythia}, \textsc{Phojet} and \textsc{Herwig/Jimmy}~\cite{herwig}, it tends to be phenomenological, requiring tuning to data. However, non-perturbative effects, such as soft diffraction, hadronization and low-\pt parton scattering, are difficult to separate experimentally. Also, the description of hard processes by the MC generators must be preserved while tuning the soft processes. The results presented here show that it is difficult to describe both MB and the UE with the same parameters.

\section{Data Samples and Selection}

The measurements presented here were made using $pp$ collision data recorded at $\sqrt{s} = 0.9$, 2.36 and 7\tev during low-luminosity running of the LHC in 2009 and the beginning of 2010. The low instantaneous luminosity ensures that there are relatively few overlapping $pp$ collisions in each bunch crossing (a background known as \textit{pile-up}), important when measuring soft-QCD observables. Data were collected with stable colliding beams and correspond to an integrated luminosity of $\int \call\,\mathrm{d}t \leq 230~\mu$b$^{-1}$. 

\subsection{Event selection\label{section:eventselection}}

Events were selected using a single-arm MBTS trigger, formed from BPTX and MBTS L1 trigger signals. The MBTS trigger was configured to require one hit above threshold on either side of the detector (2.09 $< \aeta <$ 3.84). Events were additionally required to have at least one reconstructed primary vertex, while also vetoing events with a second primary vertex with 4 or more tracks. This event selection imposes little bias on the measurements presented here, while reducing the contribution from empty events, beam background and pile-up events to a negligible level.

\subsection{Track selection}

Tracks were selected by requiring their \pt and $\eta$ to be within the specified phase space of the measurement. A good track quality was ensured by requiring a minimum number of hits in the silicon detectors, dependent on the \pt of the track. Tracks associated to particles coming from the primary interaction (known as \textit{primary}\footnote{Primary particles are defined as all particles with lifetime longer than $0.3 \times 10^{-10}$s originating from the primary interaction or from subsequent decay of particles with shorter lifetime.} particles) were selected by requiring their impact parameters, measured with respect to the reconstructed primary vertex of the event, to be within a specified range. 

\subsection{Corrections}

Data shown here have been fully corrected back to particle level in order to measure the distributions of stable (charged) particles coming from the primary $pp$ interaction. This allows a direct comparison to MC generator predictions. Corrections are applied at both the event and track level.

The event-level corrections correct for missing events due to trigger and vertex requirements. Both trigger and vertex corrections were derived from data, the former by measuring the trigger efficiency of the MBTS with respect to a control trigger using the Inner Detector and the latter by directly measuring the primary vertex reconstruction efficiency using all triggered events. 

The track-level corrections correct for detector inefficiencies and resolutions. The tracking efficiency was derived from MC samples taken through the full \textsc{Geant}~\cite{geant} detector simulation. The track-level corrections were applied in two dimensions ($\eta$, $\pt$) in order to eliminate most of the model dependence. Corrections for non-primary particles and particles outside of the kinematic range were also applied. 

All corrections were derived separately for the different analyses and phase space regions. Measurements were not extrapolated into regions of phase space not seen by the detector (e.g. very low-\pt or far-forward particles). No attempt was made to correct for the contribution coming from diffractive processes; however, phase space regions with a suppressed diffractive component were considered (see Table~\ref{table:mbphasespace}). The event selection at the particle level is always well defined and reproducible, e.g. number of charged particles $\nch \geq 2$, with $\pt > 100\mev$ and $\aeta < 2.5$. 

\section{Minimum Bias Measurements}

`Minimum bias' is an experimentally defined term, referring to the selection of inelastic events with the minimum possible requirements necessary to ensure that an inelastic collision occurred. Minimum bias events can include both non-diffractive and diffractive processes, although the precise definition and relative contributions vary among experiments and analyses. Typically, minimum bias events are dominated by soft interactions, with low transverse momentum and low particle multiplicity. 

The ATLAS MBTS trigger (see Section~\ref{section:eventselection}) is almost fully efficient, with a slightly lower efficiency in low-\nch events. In an attempt to disentangle the effects coming from diffractive processes, ATLAS has measured the properties of minimum bias events in various phase space regions~\cite{minbias}, as listed in Table~\ref{table:mbphasespace}. The high-multiplicity selections ($\nch \geq 6, \nch \geq 20$) were chosen specifically to reduce the contribution of diffractive processes to a negligible level. Kinematic properties that have been measured include the charged-particle multiplicity ($\nch$), the charged-particle transverse momentum ($\pt$) and pseudorapidity ($\eta$) spectra and the average transverse momentum of charged particles as a function of the charged-particle multiplicity ($\avgpt$ vs. \nch).

\begin{table}[htdp]
\caption{\label{table:mbphasespace} Phase space regions considered by the ATLAS minimum bias analysis~\cite{minbias}. The diffraction-suppressed phase space ($\nch \geq 6$, $\pt > 500\mev$, $\aeta < 2.5$) was used for the \textsc{Pythia} AMBT1 tune~\cite{ambt1}. The most inclusive phase space ($\nch \geq 2$, $\pt > 100\mev$, $\aeta < 2.5$) was used for the \textsc{Pythia} AMBT2b tune~\cite{ambt2}. The common LHC phase space was chosen by the LHC Minimum Bias and Underlying Event working group~\cite{lhcmbue} in order to directly compare measurements across LHC experiments.}
\begin{center}
\begin{tabular}{|c|c|c|c|c|c|c|c|}
\hline
& \multicolumn{2}{c}{Most}  & \multicolumn{2}{|c|}{Diffraction}  & \multicolumn{1}{c}{High}  & \multicolumn{2}{|c|}{LHC} \\ 
& \multicolumn{2}{c}{inclusive} & \multicolumn{2}{|c|}{suppressed}  &  \multicolumn{1}{c}{$\pt$} & \multicolumn{2}{|c|}{comparison} \\ \hline
$\nch \geq $ & 2 & 1 & 20 & 6 & 1 & 1 & 1 \\ \hline
$\pt$ [\gev] $>$ & 0.1 & 0.5 & 0.1 & 0.5 & 2.5 & 0.5 & 1.0 \\ \hline
$\aeta < $ & ~~2.5~~ & ~~2.5~~ & ~~2.5~~ & ~~2.5~~ & ~~2.5~~ & ~~0.8~~ & ~~0.8~~ \\ \hline
\end{tabular}
\end{center}
\label{default}
\end{table}%

\begin{figure}[p]
\begin{center}
\includegraphics[width=0.32\linewidth]{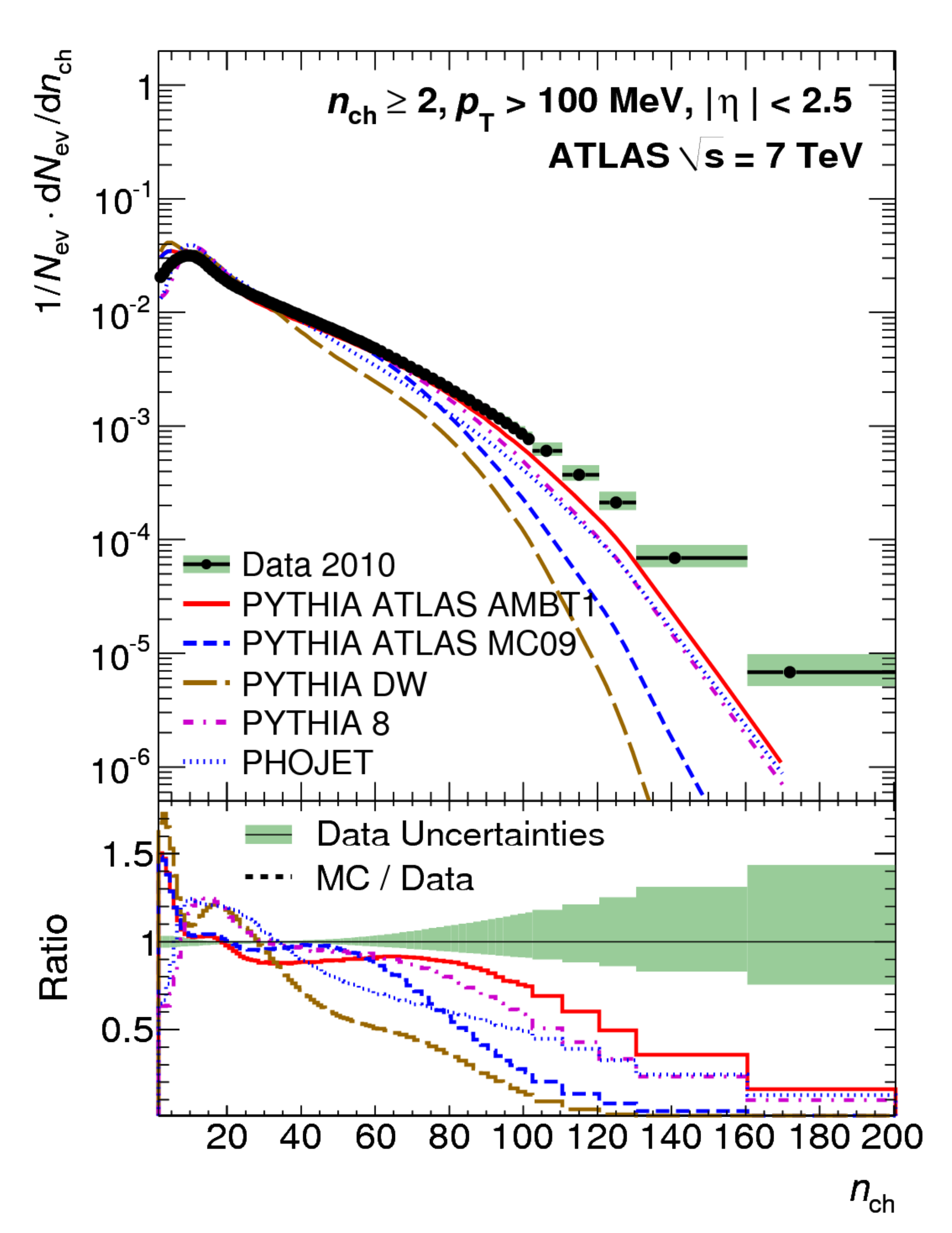}
\includegraphics[width=0.32\linewidth]{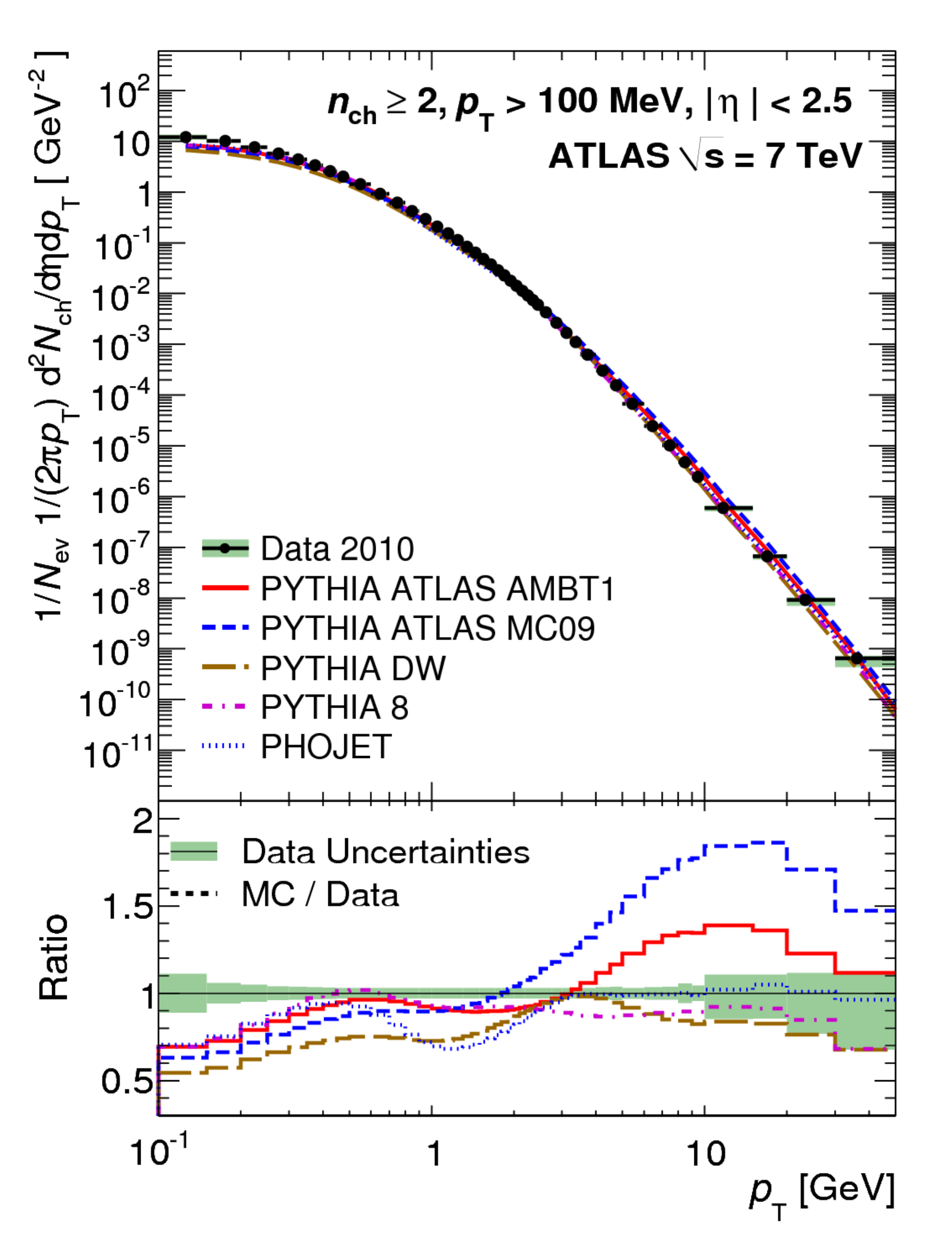}
\includegraphics[width=0.32\linewidth]{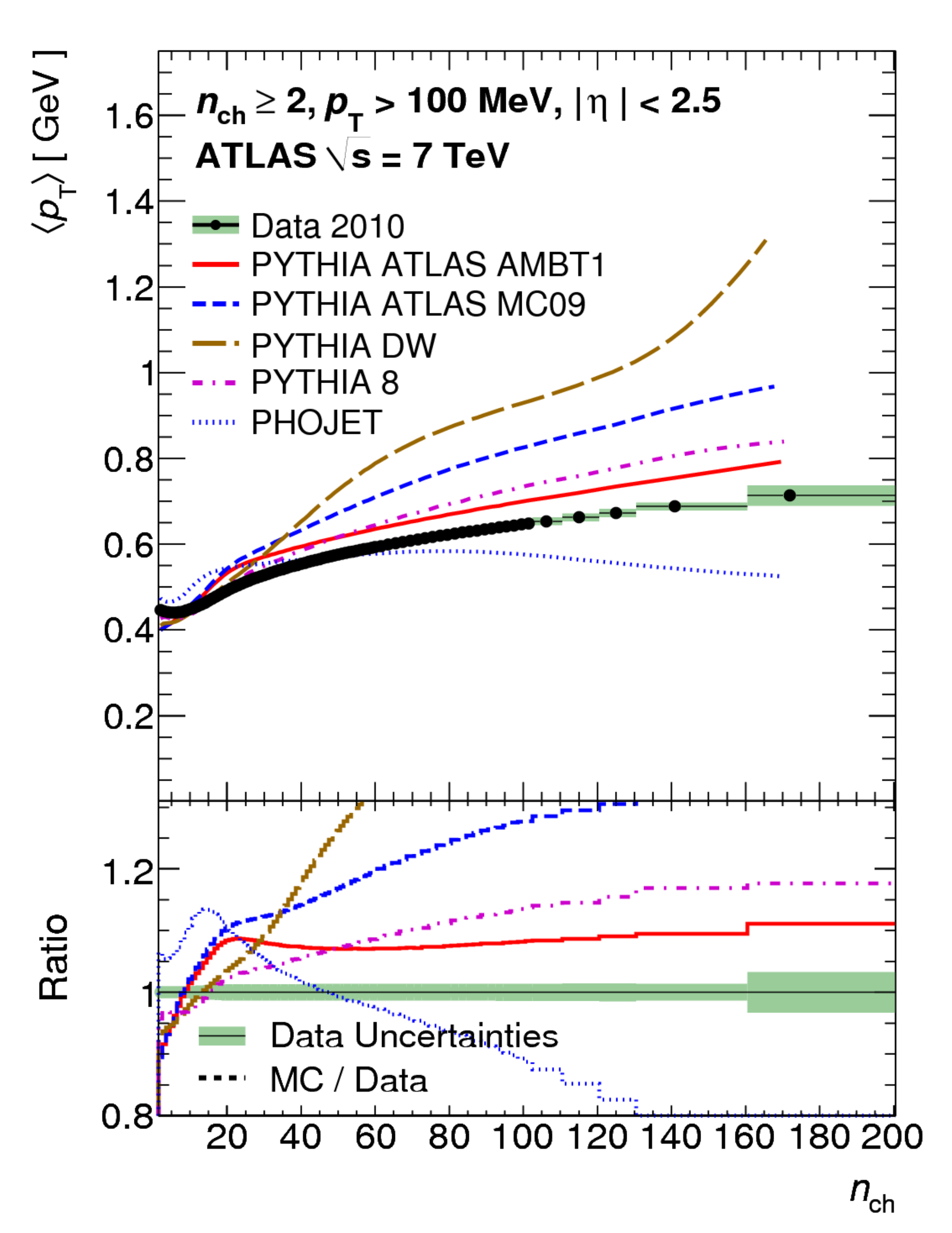}
\caption{\label{figure:mbinc}Corrected charged-particle multiplicity distribution (\textit{left}), charged-particle multiplicity as a function of transverse momentum (\textit{middle}) and the average transverse momentum as a function of the number of charged particles in the event (\textit{right}) at $\sqrt{s} = 7\tev$ for the most inclusive phase space region ($\nch \geq 2$, $\pt > 100\mev$, $\aeta < 2.5$) considered by the ATLAS minimum bias analysis~\cite{minbias}. The vertical error bars represent the statistical uncertainties, while the shaded areas show statistical and systematic uncertainties added in quadrature. The ratio of MC to data is shown at the bottom of each plot.}
\end{center}
\end{figure}

Fig.~\ref{figure:mbinc} shows the corrected charged-particle multiplicity distributions at $\sqrt{s} = 7\tev$ for the most inclusive phase space region ($\nch \geq 2$, $\pt > 100\mev$, $\aeta < 2.5$) considered by the analysis. There is an excess in the models relative to the data at low \nch and a deficiency at high \nch, a clear indication that the models have difficulty in describing both the low and high-\nch regions simultaneously. The simulation predicts a significantly harder \pt spectrum for $\pt > 3\gev$. Again here, the models have difficulty in describing the low-\pt ($\pt < 500$\mev) and high-\pt ($\pt > 3$\gev) regions simultaneously. At low values of \nch, none of the models describe the \avgpt data very well. For $\nch > 20$ the models vary widely both in slope and in absolute value. 

\begin{figure}[p]
\begin{center}
\includegraphics[width=0.32\linewidth]{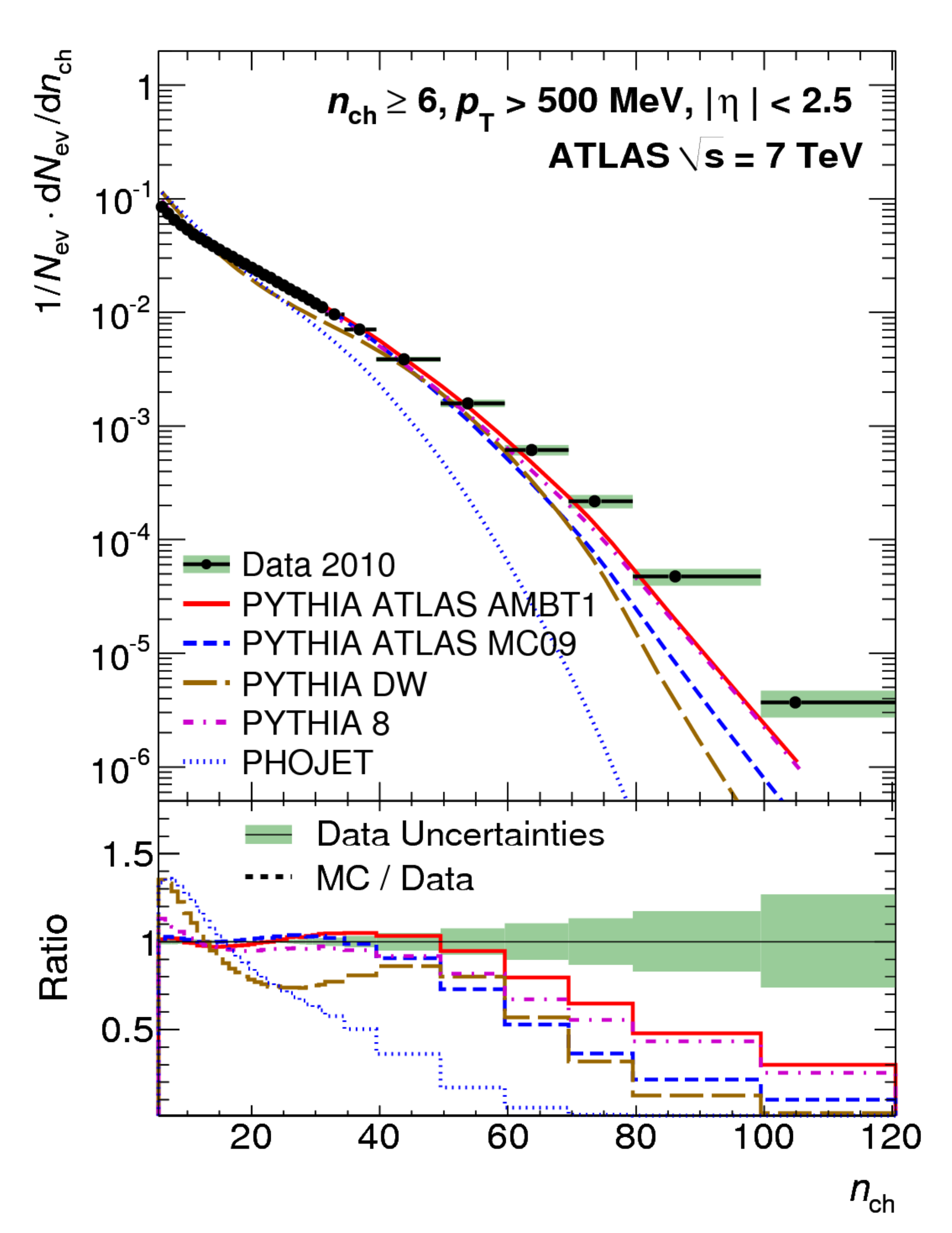}
\includegraphics[width=0.32\linewidth]{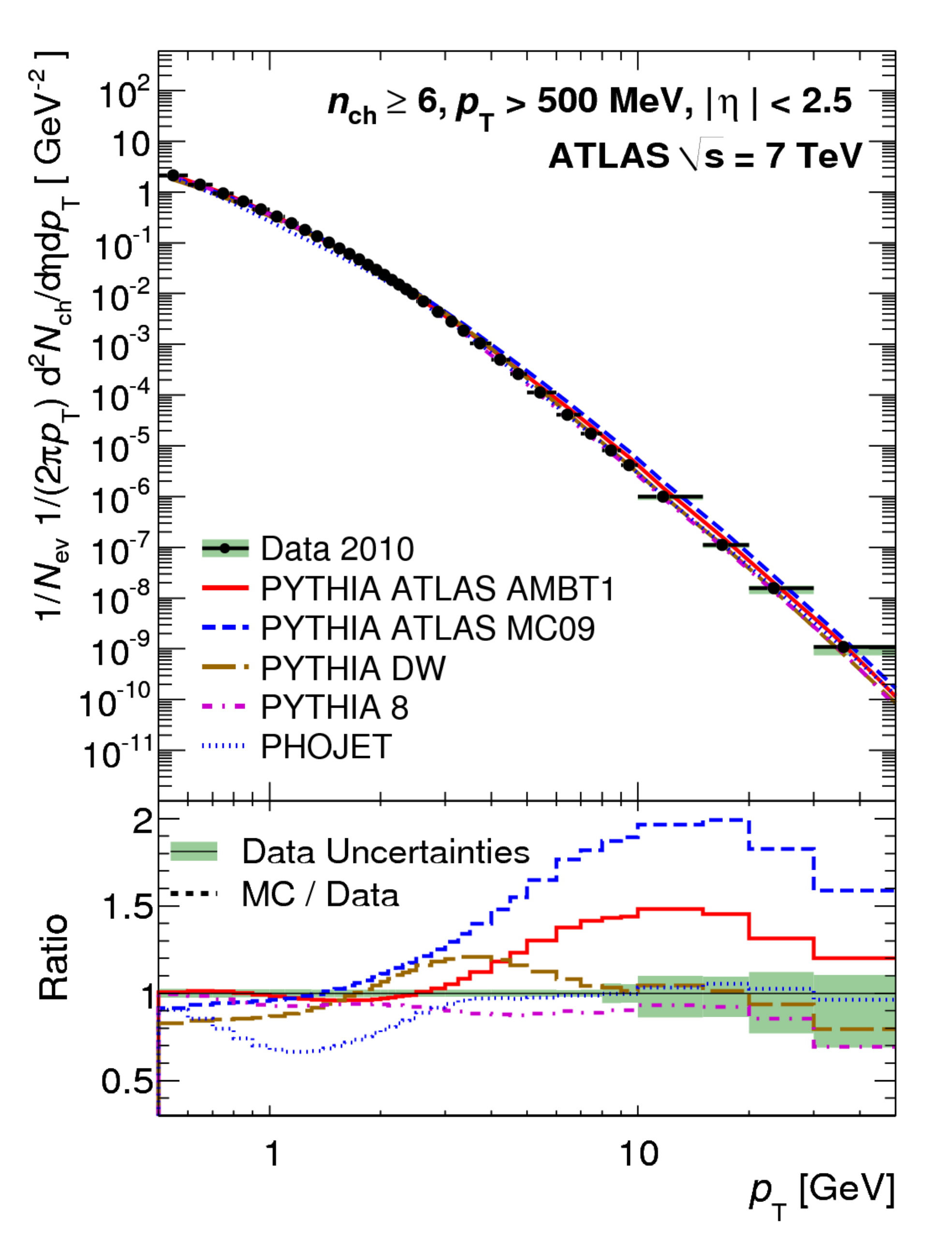}
\includegraphics[width=0.32\linewidth]{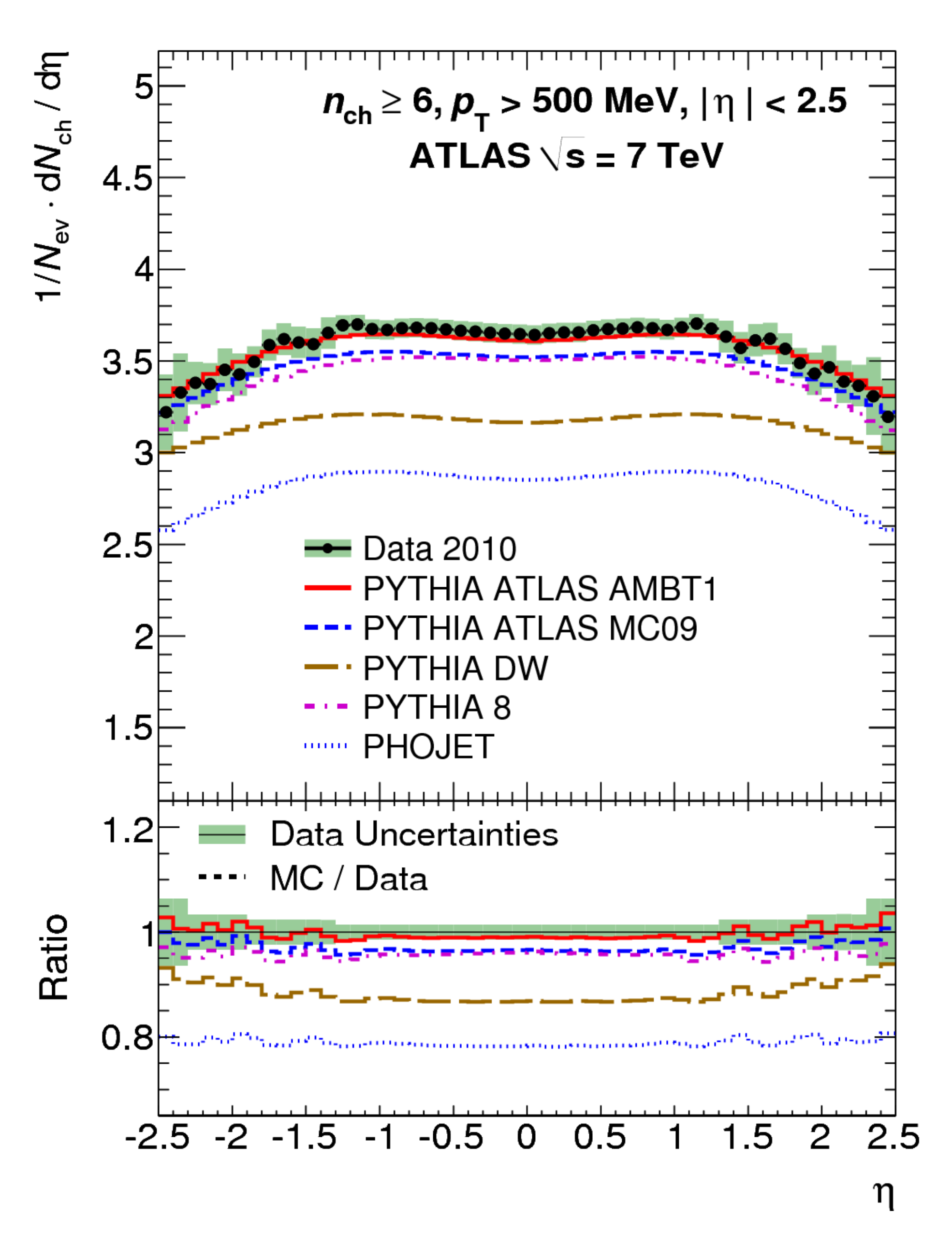}
\caption{\label{figure:mbds}Corrected charged-particle multiplicity distribution (\textit{left}) and charged-particle multiplicities as a function of transverse momentum (\textit{middle}) and pseudorapidity (\textit{right}) at $\sqrt{s} = 7\tev$ for the diffraction-suppressed phase space region ($\nch \geq 6$, $\pt > 500\mev$, $\aeta < 2.5$) considered by the ATLAS minimum bias analysis~\cite{minbias}. The vertical error bars represent the statistical uncertainties, while the shaded areas show statistical and systematic uncertainties added in quadrature. The ratio of MC to data is shown at the bottom of each plot.}
\end{center}
\end{figure}

Fig.~\ref{figure:mbds} shows the corrected charged-particle multiplicity distributions for the diffraction-suppressed phase space region ($\nch \geq 6$, $\pt > 500\mev$, $\aeta < 2.5$) at $\sqrt{s} = 7\tev$. A comparison between Figs.~\ref{figure:mbinc} and \ref{figure:mbds} shows that the modeling of diffraction plays an important role here. However, the pseudorapidity density distribution (right plot) shows that diffraction is not the only culprit for the discrepancy between data and MC.

\begin{figure}
\begin{center}
\includegraphics[width=0.5\linewidth]{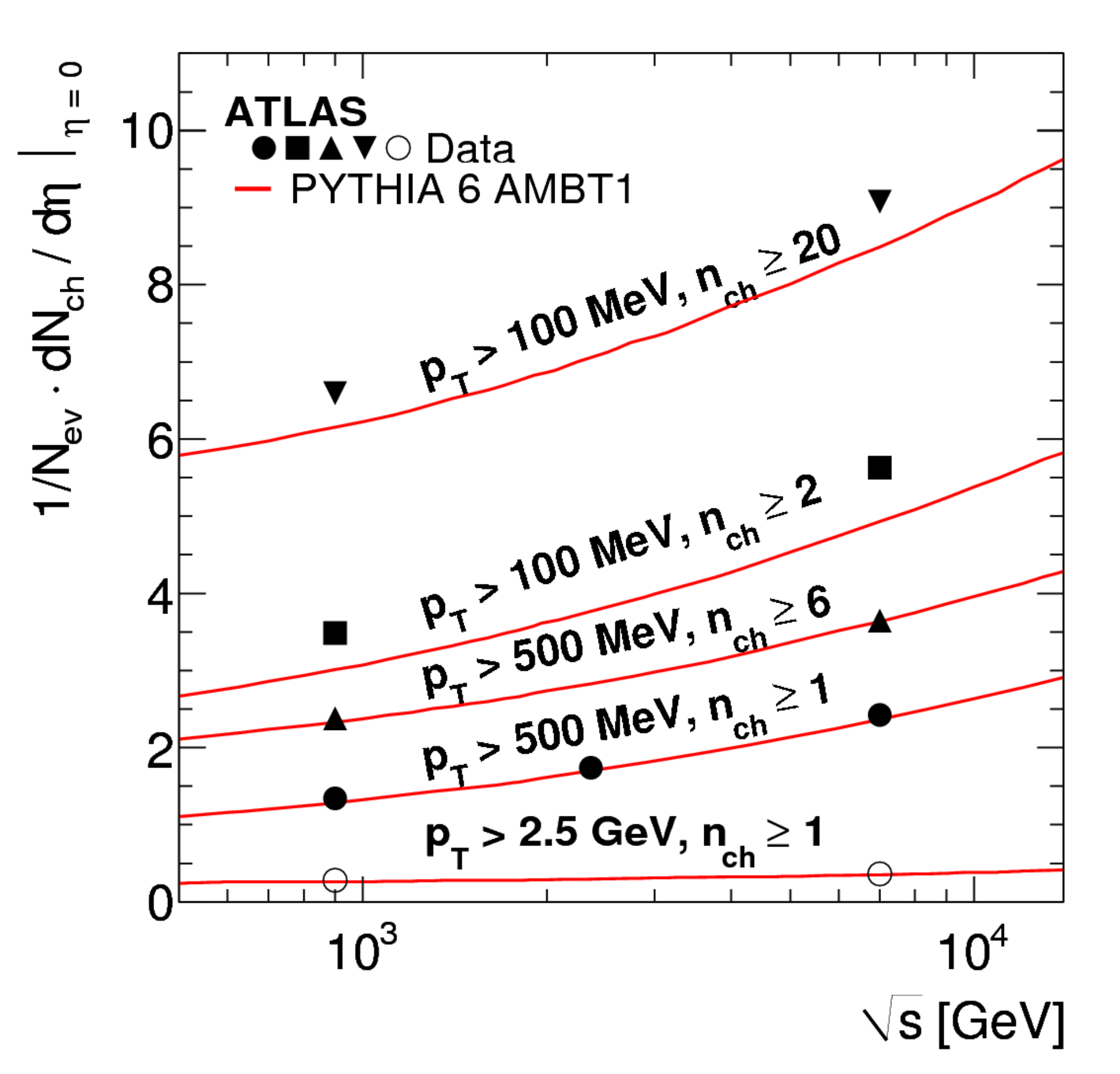}
\caption{\label{figure:mbeta_vs_s} Average charged-particle multiplicity per unit of pseudorapidity in the central region ($\eta = 0$) as a function of the center-of-mass energy $\sqrt{s}$. All measured phase space regions (see Table~\ref{table:mbphasespace}) and center-of-mass energies ($\sqrt{s} = 0.9$, 2.36 and 7\tev) from the ATLAS minimum bias analysis~\cite{minbias} are shown and compared to predictions by \textsc{Pythia} 6, with the AMBT1 tune~\cite{ambt1}. Errors shown on data points are combined statistical and systematic uncertainties.}
\end{center}
\end{figure}

The energy evolution of minimum bias properties has also been studied by measuring the same distributions at two additional center-of-mass energies from early LHC running, $\sqrt{s} = 0.9$ and 2.36\tev. Fig.~\ref{figure:mbeta_vs_s} shows a comparison of the pseudorapidity density in the central region ($\eta = 0$) as a function of $\sqrt{s}$ for all phase space regions considered by ATLAS. The data points are compared to the first ATLAS tuning of \textsc{Pythia} 6 using ATLAS data, tune AMBT1~\cite{ambt1}. The AMBT1 tune gives a good description of the energy dependence for phase spaces with $\pt > 500\mev$. However, the AMBT1 tune underestimates the amount of activity in the low-\pt region for both the most inclusive and diffraction-suppressed phase spaces. 

\section{Underlying Event Measurements}

The Underlying Event is an irreducible background to all processes at hadron colliders such as the LHC. It consists essentially of all of the soft processes accompanying a hard scatter due to additional interacting partons from the same protons. From an experimental point of view, it is impossible to separate these contributions; however, topological properties of the event can be used to define a set of physics observables that are sensitive to different aspects of the UE. 

The UE measurements performed by ATLAS thus far follow the lead of the analyses pioneered at the Tevatron by determining a high-\pt object in each event and then subdividing the azimuthal angle into \textit{toward}, \textit{transverse} and \textit{away} regions, defined according to the azimuthal angular difference $\Delta \phi$ relative to the high-\pt object. The transverse region ($60^{\circ} < |\Delta \phi| < 120^{\circ}$) is assumed to be perpendicular to the axis defined by the hard $2 \rightarrow 2$ parton process and is therefore most sensitive to activity coming from the UE. The toward region is defined $|\Delta \phi| < 60^{\circ}$ and the away region $|\Delta \phi| > 120^{\circ}$.

ATLAS has measured properties of the UE using two independent methods: a track-based measurement using tracks reconstructed by the Inner Detector and associated to charged primary particles~\cite{ue-track} and a cluster-based measurement using energy deposited in the calorimeters and associated to both charged and neutral primary particles~\cite{ue-cluster}. The activity in all three regions with respect to the leading particle (either the highest-\pt track or cluster) has been studied. Observables measured by ATLAS include the particle density, the scalar \sumpt and the average \pt of particles per event \avgpt. In both measurements, the highest-\pt particle was required to have transverse momentum $\pt^{lead} > 1\gev$.

\begin{figure}[htb]
\begin{center}
\includegraphics[height=6cm]{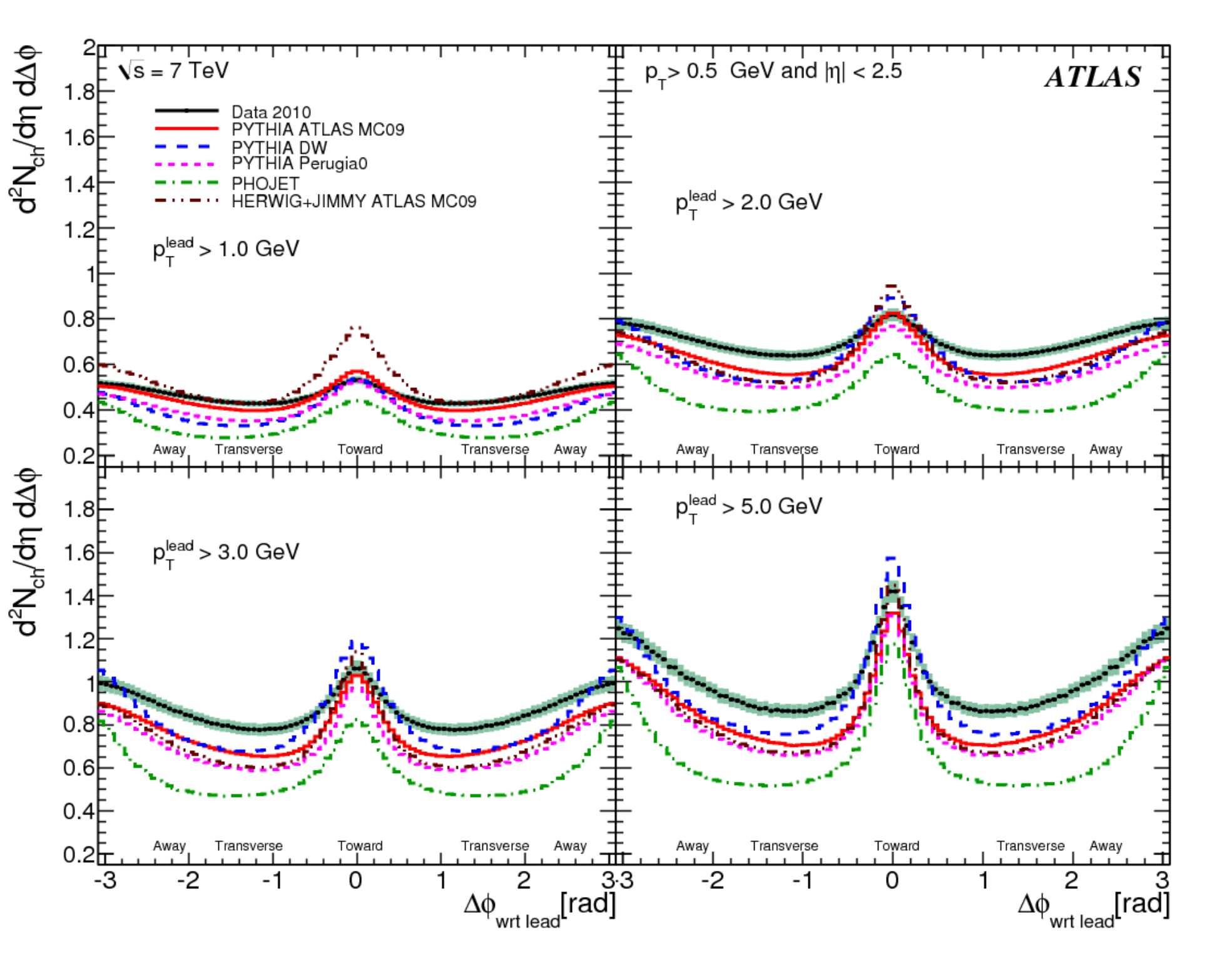}
\includegraphics[height=6cm]{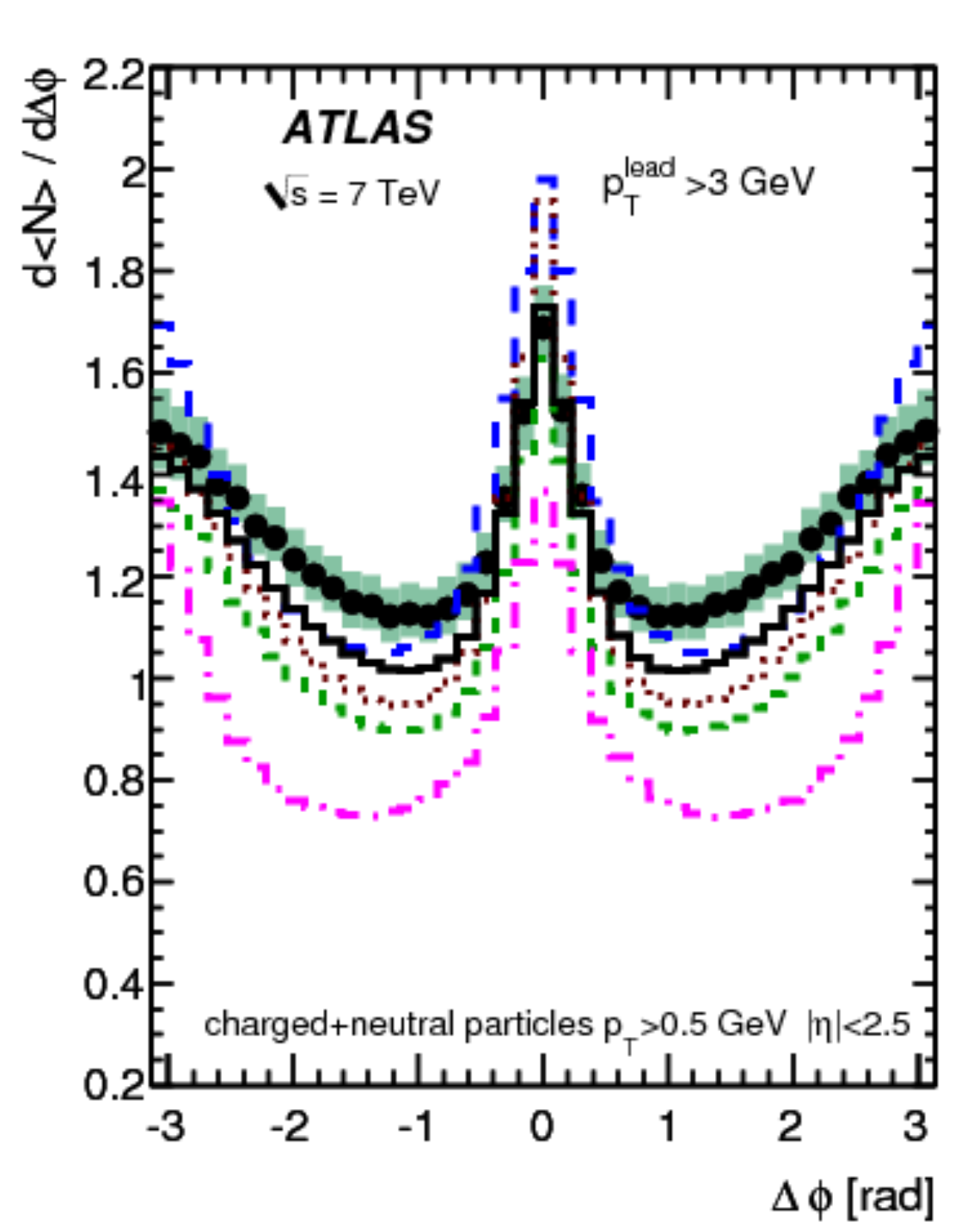}
\caption{\label{figure:ue_phi}$\phi$ distribution of (charged)-particle densities ($\pt > 500\mev$, $\aeta < 2.5$) at $\sqrt{s} = 7\tev$ with respect to the leading (charged) particle (at $\Delta \phi = 0$), from the track-based (\textit{left})~\cite{ue-track} and cluster-based (\textit{right})~\cite{ue-cluster} Underlying Event measurements by ATLAS. The leading (charged) particle has been required to have a minimum $\pt^{lead}$ as indicated on each of the plots and is excluded here. The error bars show the statistical uncertainty while the shaded areas show the combined statistical and systematic uncertainty.}
\end{center}
\end{figure}

Fig.~\ref{figure:ue_phi} shows the corrected (charged-)particle multiplicity distributions at $\sqrt{s} = 7\tev$ as a function of the azimuthal angle with respect to the leading track or particle $\Delta \phi$, for both the track- and cluster-based measurements and for various minimum values of $\pt^{lead}$. The development of a `jet-like' region of higher density in the toward and away regions is observed as the \pt of the leading track or particle increases. The amount of UE activity is underestimated by most generators by about 20\%. The particle density also has a different angular distribution than predicted by MC.

The average number of charged particles in the transverse region doubles when going from $\pt^{lead} > 2\gev$ to $> 5\gev$. Comparing the track-based measurement for $\pt^{lead} > 5\gev$ to the charged-particle distributions measured in the inclusive minimum bias spectrum, the activity in the UE is about a factor of two larger than the number of charged particles per unit pseudorapidity~\cite{ue-track}, a feature known as the `pedestal' effect. This is because the UE selection requires a greater exchange of momentum, thereby reducing the diffractive contribution. Given that there is one hard scatter in the event, it is more probable to have MPI and the UE therefore has more activity than MB.

\begin{figure}[htbp]
\begin{center}
\includegraphics[height=5.7cm]{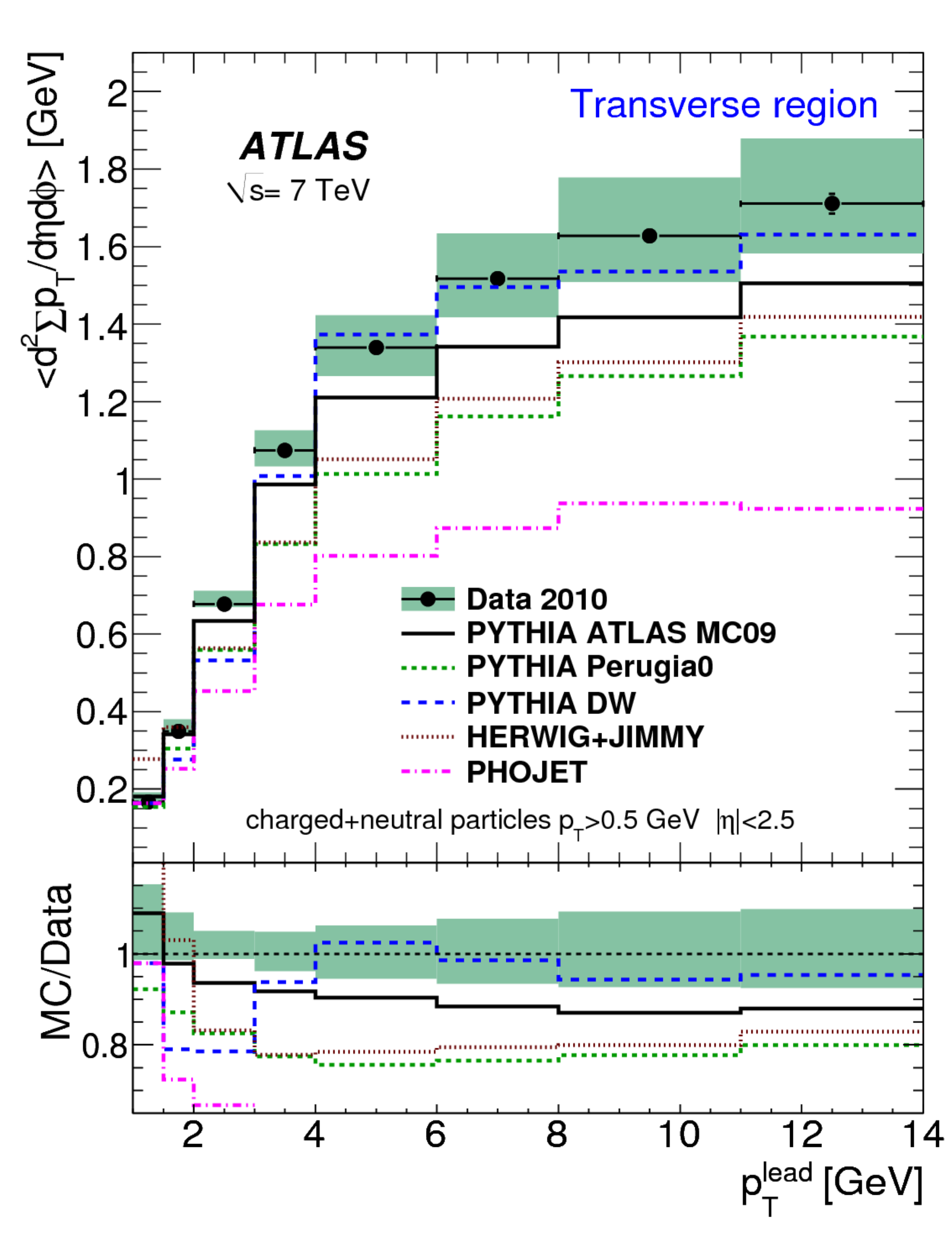}
\includegraphics[height=5.9cm]{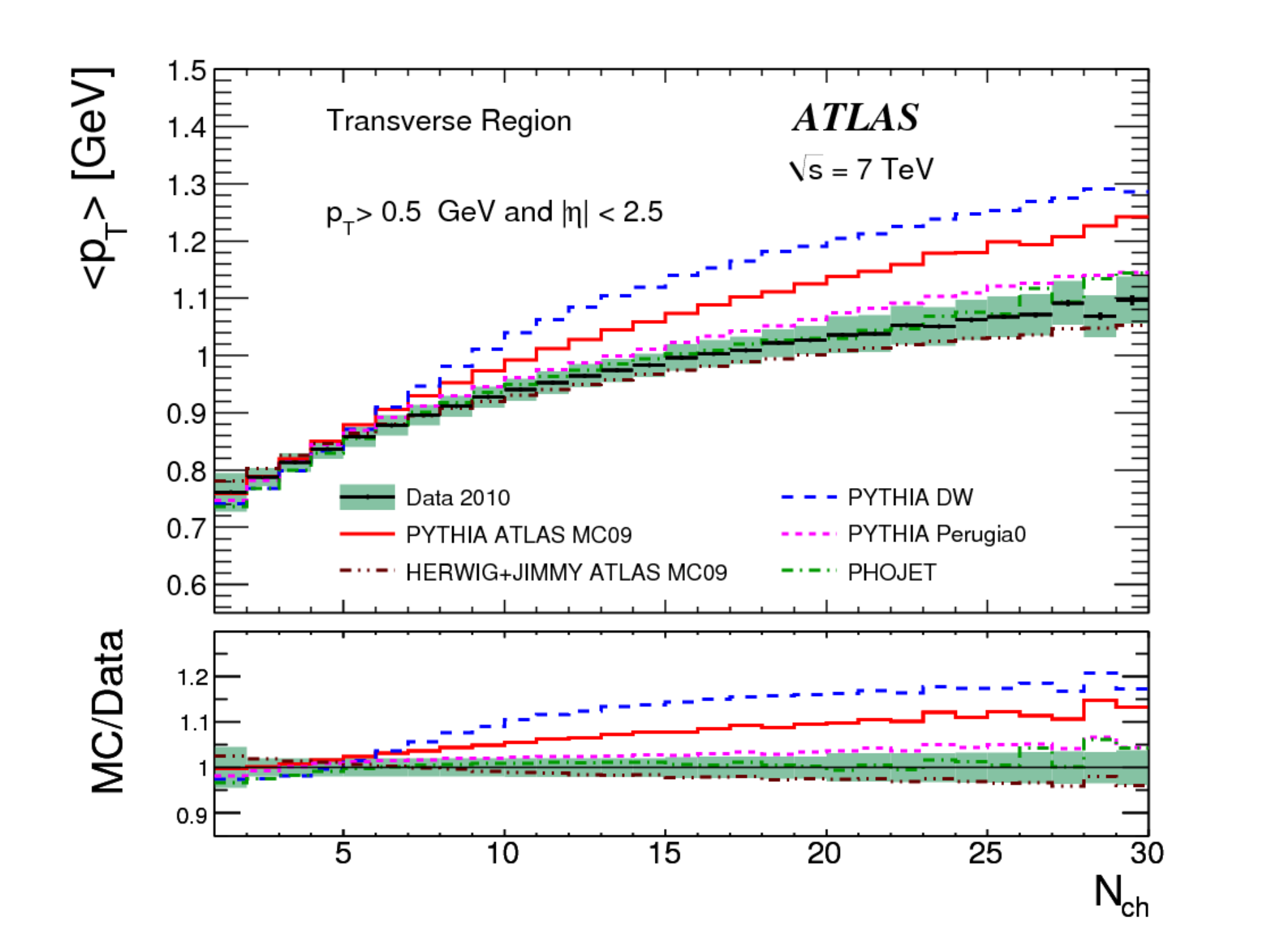}
\caption{\label{figure:ue_pt}(\textit{left}) Corrected scalar \sumpt density of stable particles ($\pt > 500\mev$, $\aeta < 2.5$) at $\sqrt{s} = 7\tev$ in the transverse region as a function of the \pt of the leading particle ($\pt^{lead}$) from the cluster-based Underlying Event measurement by ATLAS~\cite{ue-cluster}. (\textit{right}) Corrected mean \pt of charged particles at $\sqrt{s} = 7\tev$ for $\pt^{lead} > 1\gev$ as a function of the charged-particle multiplicity in the transverse region from the track-based Underlying Event measurement by ATLAS~\cite{ue-track}. For both plots, the error bars show the statistical uncertainty while the shaded areas show the combined statistical and systematic uncertainty.}
\end{center}
\end{figure}

Fig.~\ref{figure:ue_pt} (\textit{left}) shows the corrected stable particle scalar \sumpt density at $\sqrt{s} = 7\tev$ in the transverse region as a function of the \pt of the leading particle ($\pt^{lead}$) from the cluster-based measurement. The summed particle \pt in the plateau characterizes the mean contribution of the Underlying Event to jet energies. The higher number density implies a higher \pt density as well. Most of the MC tunes considered show 10-15\% lower \sumpt than the data in the plateau part of the transverse region. 

Fig.~\ref{figure:ue_pt} (\textit{right}) shows the corrected mean \pt of charged particles \avgpt at $\sqrt{s} = 7\tev$ versus the charged-particle multiplicity \nch in the transverse region from the track-based UE measurement. The correlation between \avgpt and \nch in each region is sensitive to the amount of hard versus soft processes contributing to the UE. Although not shown here, the profile in the away region is very similar to that of the transverse region, showing a monotonic increase of \avgpt with \nch. The models tend to overestimate \avgpt in both the transverse and toward regions. 

\section{Conclusions}

Data from the LHC provide a new energy scale for studying soft QCD. Charged-particle multiplicities have been measured by ATLAS in various regions of phase space, helping to disentangle the contribution coming from diffractive processes. The results of these measurements indicate a deficit of activity in models that were previously tuned to data from the Tevatron. Activity coming from the Underlying Event has been measured by ATLAS using track-based and cluster-based methods, providing statistically independent results. The activity measured in data is generally above the predictions from current model tunes. 

The (charged-)particle distributions presented here expose limitations in the phenomenological models that prevent a simultaneous description of all measured observables in all regions of phase space. This is especially true when including particles with $\pt < 500\mev$.  The AMBT1 tune of \textsc{Pythia} 6, for example, highlighted that observables such as the pseudorapidity density are not well described by MC generators, even if they were taken as input to the tuning procedure. While some of these discrepancies can be reduced by further tuning of the MC event generators, it is likely that new formulations of certain components of the models will soon be needed.

\section{Acknowledgments} 

I would like to thank the LHC and ATLAS collaborations for their contributions to these measurements. I also thank the ATLAS speakers committee and the organizing committee of the MPI workshop for the opportunity to present these results. This work was supported in part by the Alexander von Humboldt Foundation.

% %%%%%%%%%%%%%%%%%%%%%%%%%%%%%%%%%%%%%%%%%%%%%%%%%%%%%%%%%%%%%%%%%%%%%%%%%%%%%%%%

\begin{footnotesize}

\end{footnotesize}

\end{document}